\documentclass[journal = ancac3]{achemso}
\usepackage{graphicx}
\usepackage{color}
\usepackage{amsfonts}

\title{Observation of Ultrafast Free Carrier Dynamics in Single Layer MoS$_2$}
\author{Antonija Grubi\v{s}i\'{c} \v{C}abo}
\author{Jill A. Miwa}
\author{Signe S. Gr{\o}nborg}
\affiliation{Department of Physics and Astronomy, Interdisciplinary Nanoscience Center, Aarhus University,
8000 Aarhus C, Denmark}
\author{Jonathon M. Riley}
\affiliation{SUPA, School of Physics and Astronomy, University of St. Andrews,
St. Andrews, Fife KY16 9SS, United Kingdom}
\author{Jens C. Johannsen}
\affiliation{Institute of Condensed Matter Physics, \'Ecole Polytechnique F\'ed\'erale de Lausanne (EPFL), 1015 Lausanne, Switzerland}
\author{Cephise Cacho}
\author{Oliver Alexander}
\author{Richard T. Chapman}
\author{Emma Springate}
\affiliation{Central Laser Facility, STFC Rutherford Appleton Laboratory, Harwell, Didcot OX11 0QX, United Kingdom}
\author{Marco Grioni}
\affiliation{Institute of Condensed Matter Physics, \'Ecole Polytechnique F\'ed\'erale de Lausanne (EPFL), 1015 Lausanne, Switzerland}
\author{Jeppe V. Lauritsen}
\affiliation{Department of Physics and Astronomy, Interdisciplinary Nanoscience Center, Aarhus University,
8000 Aarhus C, Denmark}
\author{Philip D. C. King}
\affiliation{SUPA, School of Physics and Astronomy, University of St. Andrews,
St. Andrews, Fife KY16 9SS, United Kingdom}
\author{Philip~Hofmann}
\affiliation{Department of Physics and Astronomy, Interdisciplinary Nanoscience Center, Aarhus University,
8000 Aarhus C, Denmark}
\email{philip@phys.au.dk}
\author{S{\o}ren~Ulstrup}\altaffiliation{Present address: Advanced Light Source, E. O. Lawrence Berkeley National Laboratory, Berkeley,
California 94720, USA}
\affiliation{Department of Physics and Astronomy, Interdisciplinary Nanoscience Center, Aarhus University,
8000 Aarhus C, Denmark}
\email{sulstrup@lbl.gov}

\begin{document}
\newpage
\begin{abstract}
The dynamics of excited electrons and holes in single layer (SL) MoS$_2$ have so far been difficult to disentangle from the excitons that dominate the optical response of this material. Here, we use time- and angle-resolved photoemission spectroscopy for a SL of MoS$_2$ on a metallic substrate to directly measure the excited free carriers. This allows us to ascertain a direct quasiparticle band gap of 1.95~eV and determine an ultrafast (50~fs) extraction of excited free carriers via the metal in contact with the SL MoS$_2$. This process is of key importance for optoelectronic applications that rely on separated free carriers rather than excitons. \\
\\
KEYWORDS: transition metal dichalcogenides, MoS$_2$, free carriers, excitons, time- and angle-resolved photoemission spectroscopy.
\end{abstract}
\newpage
\maketitle

The optical properties of semiconductors are completely specified by the dynamics of free carriers and bound excitons, both stemming from electron-hole pair excitations  \cite{hopfieldtheory1958}. In bulk semiconductors such as GaAs \cite{gilleo1968} and two-dimensional (2D) quantum well systems \cite{Knox1985,robart1995}, these excitations have been studied for over half a century. Recently, genuine 2D semiconductors such as single layer (SL) MoS$_2$ have been isolated \cite{makatomically2010,splendiani2010}. In these materials the low dimensionality contributes to the presence of strongly bound excitons \cite{ramasl2012} with remarkable properties \cite{splendiani2010,Korn:2011ab,Wang:2012ag,zengvalley2012,makcontrol2012,wangvalley2013,Nie:2014aa,Ugeda2014}. 

The current understanding of the optical properties and excited carrier dynamics in SL and few layer MoS$_2$ relies largely on a series of photoluminescence and differential absorption experiments. The electron-hole pair dynamics has been investigated through the measured exciton lines, which are the most prominent features in such experiments \cite{Qiu:2013}. This approach has been remarkably successful leading to, for example, the demonstration of a valley polarization between optically induced excitons associated with the degenerate conduction band (CB) valleys in the material's band structure \cite{xiaocoupled2012,zengvalley2012,makcontrol2012}. Time-resolved studies have revealed valuable insights into exciton lifetimes, including relaxation dynamics that evolve over femto- to picosecond timescales due to scattering processes involving lattice vibrations \cite{Korn:2011ab,Wang:2012ag} and defects \cite{wangvalley2013,wangultrafast2015}. However, an important technological challenge remains: to separate the bound electron-hole pairs into free carriers that can then be extracted via metallic electrodes. This approach has recently been explored in complex 2D material heterostructures consisting of graphene and SL transition metal dichalcogenides (TMDCs) \cite{ceballosultrafast2014,hongultrafast2014}, paving the way towards efficient photovoltaic devices with improved absorption characteristics \cite{britnellstrong2013,bernardiextraordinary2013}.

\begin{figure*} [t!]
\begin{center}
\includegraphics[width=1\textwidth]{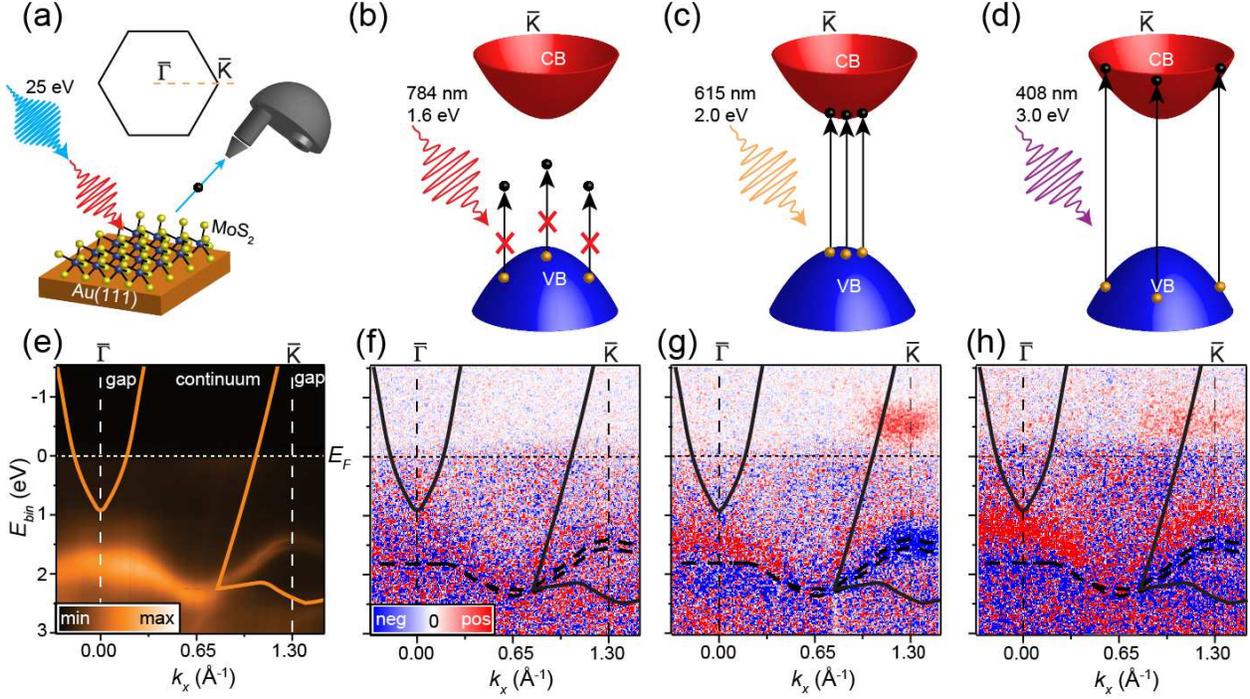}
\caption{Tuning pump-probe ARPES experiments near the direct band gap of SL MoS$_2$: (a) Schematic of our photoemission experiment. Electrons (black spheres) excited by a pump pulse tuned across the visible to infrared range are photoemitted from epitaxial SL MoS$_2$ on Au(111) using a 25~eV probe pulse and detected along the high symmetry direction $\bar{\Gamma}-\bar{\mathrm{K}}$ in the hexagonal BZ. (b)-(d) Diagrams of electron-hole pair excitations around $\bar{\mathrm{K}}$ for the studied regimes of optical excitation: (b) below the CB onset, (c) at the band gap energy and (d) above the CB onset. Holes are sketched as orange spheres. The pump pulse energies and corresponding wavelengths used in the experiments are stated. (e) Occupied band structure at equilibrium measured by synchrotron radiation. (f)-(h) Intensity difference obtained by subtracting the spectra at the peak of the optical excitation from a spectrum obtained before the arrival of the pump pulse. The orange lines in (e) and full black lines in (f)-(h) outline the boundaries between the projected bulk band continuum and gaps of Au(111). The dashed black curves in (f)-(h) are the fitted dispersion from (e). The optical excitation energies in (f)-(h) correspond to those given directly above in (b)-(d).}
\label{fig:1}
\end{center}
\end{figure*}

While the all-optical studies of SL MoS$_2$ have unveiled the exciton physics in this material, the strong exciton-related features in the spectra mask the free carrier contribution to the optical properties almost entirely \cite{Qiu:2013}. We overcome this issue by assessing the free carrier dynamics in SL MoS$_2$ with time- and angle-resolved photoemission spectroscopy (TR-ARPES). For this study, we employ a model system consisting of high quality SL MoS$_2$ epitaxially grown on a Au(111) substrate (see sketch in Fig. \ref{fig:1}(a)) \cite{miwaelectronic2015}. Not only does this system allow us to quantify the direct band gap of free carriers in a SL of MoS$_2$ on Au(111), it permits us to measure free carrier lifetimes in SL MoS$_2$ near a metal contact - a configuration that is relevant to real device architectures \cite{radisavljevic2011,mcdonnell2014}.

Using TR-ARPES to unravel the carrier dynamics near arbitrary points of a material's  Brillouin zone (BZ) has only recently become possible with the advent of high photon energy laser light via high harmonic generation \cite{Rohwer:2011,Petersen:2011,Frassetto:2011}. This is especially relevant for electronic states near the BZ boundary, such as the Dirac cone in graphene \cite{Johannsen:2013aa,Gierz:2013aa} or, indeed, the CB and valence band (VB) extrema of SL MoS$_2$, which are sketched in Fig. \ref{fig:1}(b)-\ref{fig:1}(d). In the TR-ARPES experiment, the sample is first excited by an ultrashort tunable pump pulse, such that optical pumping can be achieved with energies below (Fig. \ref{fig:1}(b)), at (Fig. \ref{fig:1}(c)) and above (Fig. \ref{fig:1}(d)) the quasiparticle band gap. The system is then probed along the $\bar{\Gamma}-\bar{\mathrm{K}}$ direction in the BZ while in the excited state by photoemission using a time-delayed 25~eV probe pulse, as sketched in Fig. \ref{fig:1}(a). Both pump and probe pulses have a duration of 30~fs.   

The electronic structure of our sample at equilibrium is characterized by sharp dispersing VB features of SL MoS$_2$ with spin-orbit split bands around the VB maximum (VBM) at $\bar{\mathrm{K}}$, as seen in the ARPES data in Fig. \ref{fig:1}(e) \cite{miwaelectronic2015}. The location of the projected bulk band continuum and bulk gaps of the underlying Au(111) have been added, because these will play a crucial role for the dynamics observed in the three different optical pumping regimes. 

The non-equilibrium carrier distribution is displayed in Fig. \ref{fig:1}(f)-\ref{fig:1}(h), which show the intensity difference between the spectrum immediately after the optical excitation (30~fs) and a spectrum taken before the arrival of the pump pulse. In these difference plots, red and blue indicate an increase and decrease of spectral intensity, respectively. As expected, such a difference spectrum is rather featureless when the system is pumped with 1.6~eV, an energy well below the band gap (Fig. \ref{fig:1}(f)). When the excitation energy is increased to 2.0~eV, we observe a clear population of the CB states at $\bar{\mathrm{K}}$, as seen in Fig. \ref{fig:1}(g). This indicates that our pump energy exceeds the band gap and excites free carriers into the CB, a point we return to below. This population strongly decreases when we excite with 3.0~eV photons (Fig. \ref{fig:1}(h)). For this excitation, one might expect a population of electronic states at higher energies in the CB and an eventual cascading down to the CB minimum (CBM). In this scenario one would either expect that the intensity at the CBM would increase at a later point in time or that the rate of decay of electrons in the CBM would slow down. We are only able to observe weak signals at the CBM with the 3.0 eV pump pulse, which we attribute to the dual constraints of phase space restrictions imposed by energy and momentum conservation rules for photo excitation of MoS$_2$ at this energy and a shorter lifetime of the MoS$_2$ states excited within the continuum of metal bulk states. The latter is fully consistent with the increase of linewidth, and thus decrease of lifetime, of the VB states as they cross into the continuum of metal bulk states, which can be seen in the equilibrium ARPES data in Fig. \ref{fig:1}(e) and in the more detailed analysis of these states in Ref. \cite{miwaelectronic2015}.

\begin{figure*} [t!]
\begin{center}
\includegraphics[width=1\textwidth]{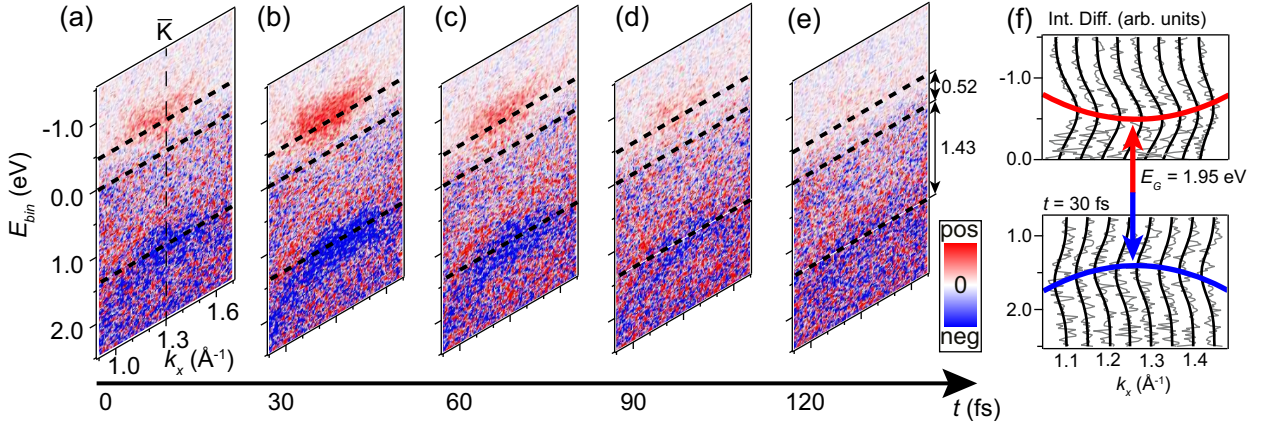}
\caption{Snapshots of excited free carriers around the $\bar{\mathrm{K}}$ point: (a)-(e) Intensity difference with respect to a spectrum obtained before optical excitation ($t<0$) at the time delays marked on the time line. Red contrast can approximately be interpreted as excited electrons while blue corresponds to excited holes. The position of the band edges and Fermi level are shown by dashed horizontal lines with the energy differences given in electron volts in (e). (f) Energy distribution curves (EDCs) of the intensity difference (grey curves) in the VB and CB regions at a time delay of 30~fs. The EDCs have been binned over a range of $\pm0.01$~\AA$^{-1}$ with respect to the given $k_x$ values. Black curves are fits to Gaussian functions. Blue and red parabolic traces follow the peak maxima in the VB and CB, respectively. The uncertainties on the band offsets in (e) and the stated band gap value ($E_G$) in (f) are $\pm0.05$~eV. All data were taken at an excitation energy of 2.0~eV.}
\label{fig:2}
\end{center}
\end{figure*}

\begin{figure} [t!]
\begin{center}
\includegraphics[width=0.38\textwidth]{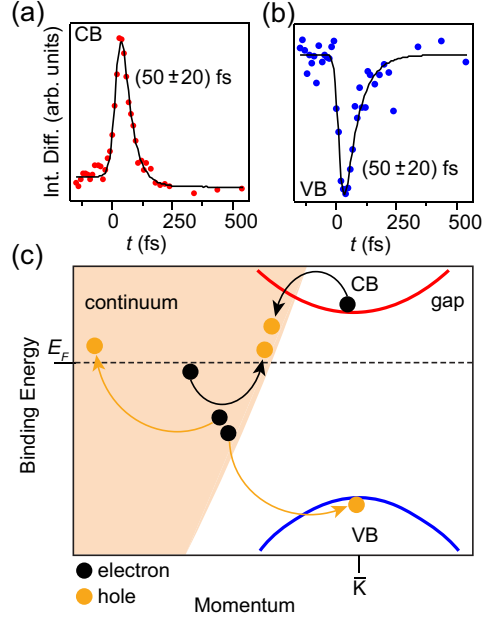}
\caption{Efficient extraction of excited free carriers by a metal contact. (a)-(b) Time dependence of the summed intensity difference (filled dots) in (a) the CB and (b) VB regions of the data. Fits to functions consisting of a rising exponential function and a single exponential decay broadened by the time resolution are shown by black lines. The stated values are the time constants of the exponential decays. (c) Interpretation of the observed dynamics. Electrons (black circles) excited into the CB of SL MoS$_2$ recombine with holes (orange circles) in the Au(111) bulk continuum (orange shaded region) by ultrafast Auger processes (black curled arrows). The same type of processes (orange curled arrows) are responsible for the recombination of holes in the VB of MoS$_2$ with electrons in Au(111).}
\label{fig:4}
\end{center}
\end{figure}

The data in Fig. \ref{fig:1}(g) establish that optical pumping with 2.0~eV photons is required to resolve the CBM around $\bar{\mathrm{K}}$ and thus to observe the free carrier dynamics of SL MoS$_2$. We have therefore used this excitation energy while recording the time-resolved series of snapshots of this region of the BZ. Snapshots of this series are shown in  Fig. \ref{fig:2} while the quantitative intensity development is discussed later in connection with Fig. \ref{fig:4}. Excited electrons and holes are already distributed around the VB and CB edges during the optical excitation phase at $t=0$, i.e. at the centre of the pump pulse (see  Fig. \ref{fig:2}(a)). This is ascribed to rapid carrier-carrier scattering. The peak excitation is reached at $t=30$~fs  in Fig. \ref{fig:2}(b) (see also Fig. \ref{fig:4}), where the full parabolic dispersions of the band extrema emerge due to the increased population of carriers. This then rapidly relaxes with the overall intensity difference already halved at $t=60$~fs in Fig. \ref{fig:2}(c) and approaching the noise level at $t=90$~fs in Fig. \ref{fig:2}(d). In Fig. \ref{fig:2}(e), at $t=120$~fs, no excited carriers can be seen. 

The parabolic dispersion around the VB and CB edges can be extracted from the peak intensity difference at $t=30$~fs using the analysis introduced in Fig. \ref{fig:2}(f). This allows us to extract values for the VB and CB offsets with respect to the Fermi level, as indicated in Fig. \ref{fig:2}(e). We can therefore directly measure a free carrier direct quasiparticle band gap of ($1.95 \pm 0.05$)~eV, which is marked in Fig. \ref{fig:2}(f). This value is different from a previous estimate by high resolution ARPES using the standard approach of populating the CB by alkali-doping \cite{miwaelectronic2015,zhangdirect2014}. We ascribe this to the possibility of a severe band distortion caused by alkali-MoS$_2$ bonding \cite{Komesu:2014ab,miwaelectronic2015}, inhomogeneous doping or band tailing effects.  The observed gap is substantially smaller than a theoretically estimated value of 2.8~eV for free-standing MoS$_2$ \cite{Qiu:2013}. This can be explained by a strong renormalization of the band gap due to the metallic substrate. Indeed, similar renormalization effects, although substantially weaker, have already been observed for 2D TMDCs grown on graphene on silicon carbide and graphite \cite{zhangdirect2014,Ugeda2014}. Note that the quasiparticle gap that we determine for excited free carriers here is a fundamentally different quantity from the optical gap that derives from the energy of the A exciton line. The size of the optical gap is around 1.9~eV for exfoliated MoS$_2$ placed on insulating substrates such as SiO$_2$/Si\cite{zengvalley2012} and hBN/SiO$_2$/Si\cite{makcontrol2012}. In our case, the close proximity of SL MoS$_2$ to the underlying metallic substrate is expected to cause a significant screening of the excitons\cite{Ugeda2014}. This reduces the exciton binding energy, which may explain why the measured size of the quasiparticle gap for MoS$_2$ on Au(111) nearly coincides with the optical gap for MoS$_2$ on SiO$_2$/Si.

\begin{figure} [t!]
\begin{center}
\includegraphics[width=0.38\textwidth]{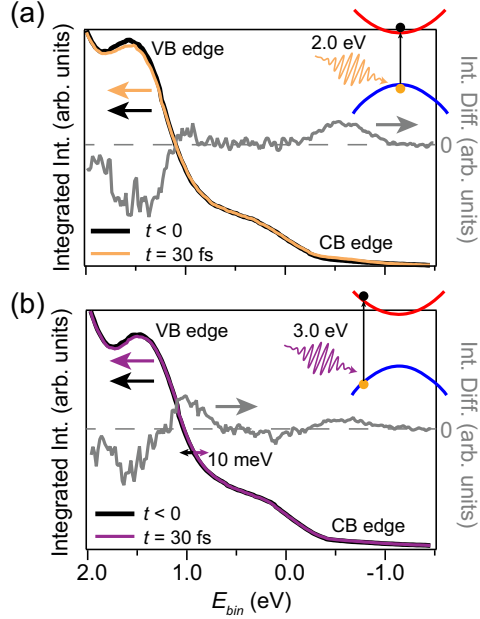}
\caption{Analysis of the photoemission intensity for excitations at the gap energy (2.0~eV) and above (3.0~eV): (a) EDCs at 2.0~eV optical pumping, extracted by integrating the absolute photoemission intensity over $\pm0.3$~\AA$^{-1}$ with respect to $\bar{\mathrm{K}}$. An equilibrium EDC obtained before optical excitation ($t<0$) and an EDC at the peak excitation ($t=30$~fs) are shown for comparison. The CB (VB) edge is identified via a small surplus (depletion) of intensity, which can also be seen by a peak (dip) in the plotted difference between the two EDCs (grey curve). The excitation process around the VB and CB edges is defined in the upper right diagram. (b) Plots of EDCs integrated over $\pm0.3$~\AA$^{-1}$ with respect to $\bar{\mathrm{K}}$ as in (a) but at a photon energy of 3.0~eV, well above the gap energy (see upper right diagram for definition of this process). A shift on the order of 10~meV between the two EDCs is highlighted by a double-headed arrow. This is clearly visible via an increase in the difference curve around a binding energy of 1.0~eV.}
\label{fig:3}
\end{center}
\end{figure}

A more detailed analysis of the photoemission intensity for a pump excitation with 2.0~eV photons is presented in Fig. \ref{fig:3}(a). The black and orange curves are energy distribution curves (EDCs) of the photoemission intensity at $t<0$ and $t=30$~fs, respectively, integrated over the momentum range that contains the CB and VB edges. The EDC at $t<0$ was obtained from a spectrum acquired 200~fs before the arrival of the optical excitation. Pumping the system causes the $t=30$~fs spectrum to be slightly different from the equilibrium situation at $t<0$, with a depletion of intensity in the VB part and a small rise of intensity above the noise level in the CB region. This enhanced population of electrons around the CB edge and the corresponding holes around the VB edge are clearly seen in the difference between these EDCs (grey curve in Fig. \ref{fig:3}(a)) in a similar way as in the corresponding Fig. \ref{fig:1}(g). The same analysis for the 3.0~eV optical pumping regime is shown in Fig. \ref{fig:3}(b). The analysis reveals that the intensity difference signal in the VB region is instead dominated by a small shift of the EDC at $t=30$~fs to lower binding energies, compared to the EDC taken before excitation. This yields a substantial increase (decrease) in the difference curve on the lower (higher) binding energy side of the VB peak, which is also visible in the entire MoS$_2$ VB in Fig. \ref{fig:1}(h). While the changes in the intensity difference seem dramatic, this shift is merely on the order of 10~meV as seen in the EDCs in Fig. \ref{fig:3}(b). We can rule out that the shift is caused by a surface photovoltage effect or by a photoelectron space charge above the sample surface because only the the MoS$_2$ states shift. Both aforementioned effects would have an impact on the photoelectron propagation in vacuum and thereby lead to shifts of both MoS$_2$ and Au states \cite{Yang:2014aa,Ulstrup:2015j}.

For the data taken with a 2.0~eV pump pulse, the shift appears insignificant in the intensity difference, and we can reliably assume that the time dependence of the intensity in the CB and VB regions originates entirely from the electron-hole pair dynamics induced by the pump pulse. This is studied in further detail in Fig. \ref{fig:4}(a)-\ref{fig:4}(b), showing the time-dependent integrated CB and VB intensity, respectively. In both cases we find the decay part to be well-described by a single exponential function with a time constant of 50~fs, which is at the limit of the time-resolution of our experiment. The free carrier dynamics observed here are several orders of magnitude faster than the dynamics reported for bulk GaAs which is also a direct band gap semiconductor \cite{Yang:2014aa}. The 50~fs time-scale is too short for lattice excitations to take place.  The observed ultrafast dynamics can therefore only be explained by Auger-type electron-electron interactions directly involving the metal contact, as sketched in Fig. \ref{fig:4}(c). Excited electrons in the CBM of SL MoS$_2$ can easily find a hole in the nearby bulk continuum of states in the metal while simultaneously exciting an electron-hole pair in the Fermi sea of the metal. The reverse of these processes accounts for the hole dynamics in the VB.

Finally we note that the free carrier dynamics we have observed here are also much faster than any other free carrier decay mechanisms - for example, exciton condensation, recombination in trap states originating from defects or radiative recombination. This suggests that extremely efficient free carrier extraction is already present near gold electrodes in contact with the SL MoS$_2$. This type of material combination thus comprises a route towards exploiting the optically generated free carriers in novel 2D semiconductor systems.  \\
 \\

\textbf{Methods.} The SL epitaxial MoS$_2$ samples were grown by a physical vapor deposition (PVD) procedure consisting of Mo evaporation on a clean single crystal Au(111) substrate in an ultra-high vacuum (UHV) chamber with a low pressure of H$_2$S gas, followed by annealing in the H$_2$S atmosphere. The growth was cycled to provide large-area SL MoS$_2$ as described in further details in Ref. \cite{sorensenstructure2015}. High resolution equilibrium ARPES data were obtained by transferring the sample through air to the SGM-3 UHV end-station of the synchrotron radiation source ASTRID2 in Aarhus, Denmark. The sample was annealed to 500~K to remove adsorbed species and to give rise to sharp MoS$_2$ electronic states in ARPES spectra (see Fig. \ref{fig:1}(e)). The ARPES data were collected with the sample temperature kept at 70~K. A photon energy of 25~eV was selected in order to be consistent with the TR-ARPES data. The total energy- and angular-resolution amounted to 20~meV and 0.2~$^{\circ}$, respectively.

The TR-ARPES data were acquired at the Artemis facility, Rutherford Appleton Laboratory in Harwell, UK \cite{Frassetto:2011}. Samples were transported in an evacuated tube pumped down below $1\cdot10^{-9}$~mbar. Once the samples were placed in the TR-ARPES end-station they were annealed to 500~K to remove any adsorbed surface contaminants. The sample temperature was kept at 50~K using a liquid helium cryostat during measurements.  A 1~kHz Ti:sapphire amplified laser system with a wavelength of 785~nm, a pulse duration of 30~fs and an energy per pulse of 12~mJ was used to generate pump and probe pulses. In order to measure the band structure throughout the entire first BZ of SL MoS$_2$, probe pulses of $h\nu=25$~eV, corresponding to the 15th harmonic of the laser fundamental, were generated by focusing a part of the laser energy on a pulsed jet of argon gas. The remaining laser energy was used to drive an optical parametric amplifier (HE-Topas) followed by a frequency mixing stage, which enabled us to obtain tunable pump pulses with wavelengths centered at 408~nm, 615~nm and 784~nm. The applied pump fluences were kept in the range 1.4 to 4.8~mJ/cm$^2$, and were always optimized to avoid pump-induced space-charge broadening of the acquired spectra \cite{Ulstrup:2015j}. We used a configuration where the pump pulse was $s$-polarized and the probe pulse was $p$-polarized. The time delays of pump and probe pulses were varied using a mechanical delay line. In order to achieve a satisfactory signal to noise ratio in the pump-probe spectra total acquisition times on the order of 17~hours per dataset were necessary. The energy-, angular- and time-resolution were 400~meV, 0.3$^{\circ}$ and 40~fs, respectively. The intensity difference in Fig. \ref{fig:4}(a) was summed over a binding energy range from $-0.95$~eV to $-0.20$~eV and a momentum range from $1.0$~\AA$^{-1}$ to $1.5$~\AA$^{-1}$. In Fig. \ref{fig:4}(b) the sum was performed over the same momentum range but over a binding energy range from $1.00$~eV to $1.75$~eV. The exponential function fits of these data were performed using a Gaussian broadening with a width of 40~fs. Our terminology is such that EDCs are energy distributions at fixed $k_{\parallel}$, not at fixed emission angle.

\section{Acknowledgements}
We thank Phil Rice for technical support during the Artemis beamtime. We gratefully acknowledge funding from the VILLUM foundation, the Lundbeck foundation, Haldor Tops{\o}e A/S, the Danish Strategic Research Council (CAT-C), EPSRC (Grant Nos. EP/I031014/1 and EP/L505079/1), The Royal Society and the Swiss National Science Foundation (NSF). Ph. H. and S. U. acknowledge financial support from the Danish Council for Independent Research, Natural Sciences under the Sapere Aude program (Grant Nos. DFF-4002-00029 and DFF-4090-00125). Access to the Artemis Facility was funded by STFC. Open-access data underpinning this publication can be accessed at\\ \url{http://dx.doi.org/10.17630/35bf1069-6ed4-4b08-9eca-0bf5483ad753}.

\newpage
￼\makeatletter \renewcommand\@biblabel[1]{(#1)}
\makeatother
%\bibliography{femtorefs}

\end{document}